\documentclass[a4paper,11pt]{article}

\usepackage{a4}
\usepackage{jheppub}
\usepackage{graphics}
\usepackage{amssymb,amsmath}
\usepackage{color}
\usepackage{axodraw}
\usepackage{wasysym}
\usepackage[tight]{subfigure}
\usepackage{booktabs}

\newcommand{\neu}[1]{\tilde{\chi}^0_{#1}}
\newcommand{\cha}[1]{\tilde{\chi}^\pm_{#1}}

\newcommand{\gev}{\ \mathrm{GeV}}

\def\gsim  {\hspace{0.3em}\raisebox{0.4ex}{$>$}\hspace{-0.75em}\raisebox{-.7ex}{$\sim$}\hspace{0.3em}}

\preprint{DESY-12-158}
\title{Can R-parity violation hide vanilla supersymmetry at the LHC? }
\author[a]{Masaki Asano,}
\author[b,1]{Krzysztof Rolbiecki,%
\note{From 1st October 2012 at the Instituto de F\'{\i}sica Te\'{o}rica, IFT-UAM/CSIC, 28049 Madrid, Spain.}}
\author[b]{and Kazuki Sakurai}
\affiliation[a]{
II. Institute for Theoretical Physics, University of Hamburg, 
Luruper Chausse 149, DE-22761 Hamburg, Germany}
\affiliation[b]{DESY, Notkestrasse 85, D-22607 Hamburg, Germany}
\emailAdd{masaki.asano@desy.de}
\emailAdd{krzysztof.rolbiecki@desy.de}
\emailAdd{kazuki.sakurai@desy.de}
\abstract{
Current experimental constraints on a large parameter space in supersymmetric 
models rely on the large missing energy signature. This is usually provided by the lightest 
neutralino which stability is ensured by $R$-parity. However, if 
$R$-parity is violated, the lightest neutralino decays into the standard model 
particles and the missing energy cut is not efficient anymore. In particular, 
the $UDD$ type $R$-parity violation induces the neutralino decay to three quarks 
which potentially leads to the most difficult signal to be searched at hadron colliders.  
In this paper, we study the constraints on $R$-parity violating supersymmetric 
models using a same-sign dilepton and a multijet signatures. We show that the gluino and squarks lighter than TeV  
are already excluded in the constrained minimal supersymmetric standard model with the $R$-parity violation if 
their masses are approximately equal. We also analyze constraints in a simplified model with the $R$-parity violation. 
We compare how the $R$-parity violation changes some of the observables typically used to distinguish a supersymmetric signal from standard model backgrounds. 
}

\begin{document}
\maketitle
\flushbottom


\section{Introduction\label{intro}}

The LHC experiments are searching for a new physics which could naturally explain the mechanism of 
electroweak symmetry breaking. One of the leading candidates 
for the new physics is the low energy supersymmetry (SUSY). In the minimal 
supersymmetric standard model (MSSM), the electroweak symmetry breaking 
scale is determined by the SUSY breaking mass parameters and the $\mu$ term. Then, 
to achieve the correct electroweak symmetry breaking without fine-tuning, 
SUSY particle masses would have to be of the same order as the electroweak scale. 
However, ATLAS and CMS have already excluded gluino and squarks 
with equal masses below 1.4 TeV in the constrained MSSM (CMSSM) with 
$R$-parity~\cite{:2012rz,:2012mfa}. In particular, the large 
missing energy signature plays an important role in searches for the 
SUSY signal. 

If $R$-parity is violated, the missing energy distribution changes 
drastically because the lightest SUSY particle (LSP) is no longer stable. 
The amount of missing energy decreases and discrimination from the standard 
model (SM) background becomes tricky. Thus, the constraints on SUSY 
particle masses would relax in such a case. Especially, in the case when the 
neutralino LSP decays into three quarks via the $UDD$-type 
$R$-parity violating coupling the search for SUSY signal  can be challenging at hadron colliders because charged leptons and missing energy 
are not produced in the LSP decay~\cite{:1999ATLASTDR}.

Existing studies of the $UDD$-type $R$-parity violation focus on rather specific spectra and processes,
e.g.\ stop LSP~\cite{Allanach:2012vj}, gluino LSP~\cite{Chatrchyan:2011cj}, 
stop direct production processes with stop and neutralino LSP~\cite{Evans:2012fu}
and gravitino LSP with $\tilde q \to \tilde \chi_1^0 j \to \tilde \ell \ell j \to \ell \ell j \tilde G$~\cite{Chatrchyan:2012ye}. 
On the other hand, the aim of this paper is to see the impact of the $UDD$-type $R$-parity violation on the LHC 
constraints on the ``vanilla''-type SUSY spectrum
(with neutralino LSP, a GUT relation in gaugino masses and non-decoupling first and second generation squarks). 
This is a 
first step to understand whether $R$-parity violation helps to hide the low energy SUSY spectrum below 1~TeV.

In this paper, we show that SUSY signals with this type of  $R$-parity violation 
are already strongly constrained by current searches. We reinterpret the CMS analysis 
of same-sign dilepton signal~\cite{:2012th}.
Such a signal can be expected in the low energy SUSY because of superpartners charged under $SU(2)$ that would subsequently  
decay into a weak boson and other superpartner. In such a case, leptons 
from the weak boson decay would appear in the final state. Furthermore, we also 
investigate the constraints from the ATLAS search of final states with a
large jet multiplicity and missing transverse momentum~\cite{Aad:2012hm}. This is motivated by the $UDD$ coupling which leads to the neutralino decay to three jets. Hence,  
one would expect an increased number of jets in the final state compared to usual benchmark scenarios. Neutrinos from gauge boson decays and jet energy mismeasurement
would provide required amount of missing energy to pass a cut.
For the purpose of this analysis, we investigate the CMSSM and a 
simplified model as benchmark spectra. 
We show that these ATLAS and CMS searches already provide a  
good sensitivity. Gluino and squarks with equal masses below 900 GeV 
are excluded in the CMSSM with the $UDD$ operator. 

The paper is organized as follows. In section~\ref{sec:2} we discuss general properties of $R$-parity violating signal at the LHC. In section~\ref{sec:3} we define signal regions and specify details of event generation. Section~\ref{sec:4} discusses the constraints on a CMSSM model and a simplified model  with $R$-parity violation derived from LHC searches introduced in section~\ref{sec:3}.  Finally we conclude in section~\ref{summary}.  


\section{R-parity violating SUSY signal \label{sec:2}}

We consider the following $R$-parity violating term in
the superpotential,\footnote{For a general review of $R$-parity violating SUSY models, see for instance ref.~\cite{Barbier:2004ez}.}
\begin{eqnarray}
W \supset
   \lambda^{''}_{212} U_2 D_1 D_2 
    \qquad {\rm with } \qquad
    \lambda^{''}_{212} \sim 1 \times 10^{-3},
   \label{eq:UDD}
\end{eqnarray}
where $U_i, D_j$ are generation $i, j$ right-handed quark superfields.
In the analyzed models the lightest supersymmetric particle will be the lightest neutralino. The lightest neutralino subsequently decays as
\begin{equation}
\neu{1} \to c\;d\;s  \qquad \mathrm{or} \qquad \neu{1} \to \bar{c}\;\bar{d}\;\bar{s} \label{eq:decay}
\end{equation}
via the $R$-parity violating interaction in a 3-body decay with an off-shell squark.
Decay pattern of the other SUSY particles remains almost unaffected compared to the $R$-parity conserving scenario.

The neutralino cannot be a dark matter candidate since it is no longer stable on cosmological time scales.
The dark matter in the Universe could be explained by non-SUSY particles, such as an axion, in this setup.
It is known that introduction of a large $R$-parity violating (RPV) coupling would wash out the baryon asymmetry
before the electroweak phase transition.
The bounds on the RPV couplings are roughly ${\cal O} (10^{-7})$~\cite{Campbell:1990fa, Fischler:1990gn, Dreiner:1992vm}.
However, this constraint can be easily avoided in scenarios where the baryon asymmetry is generated
after the electroweak transition, as in the electroweak baryogenesis and in Affleck-Dine baryogenesis with a long lived
condensate or Q-ball.    
The proton lifetime does not constrain our models, since the lepton number is still conserved.
The $n - \bar n$ oscillation constraint is also satisfied since only one of the three flavour indices in eq.~\eqref{eq:UDD}
involves the first generation.

We choose the coupling eq.~\eqref{eq:UDD} and the neutralino LSP as potentially the most difficult case
to be searched at the LHC.
The $U_2 D_1 D_2$ operator does not produce charged leptons or $b$-jets in the neutralino decay.
On the other hand, if the RPV operators involving third generation are introduced,
bottom quarks and $W$ bosons would be more abundant, making the search much easier~\cite{Choudhury:2005dg,Bomark:2011ye,AbdusSalam:2011fc,Csaki:2011ge,Allanach:2012vj,Brust:2012uf}.
Furthermore, in a slepton LSP case, the constraints may be stronger than in
the neutralino LSP case, because of multi-lepton final states. 
The available parameter space where a chargino is the LSP is not large~\cite{Giudice:1995qk} and the signal would be rather similar to the neutralino LSP case.

In our scenarios, the neutralino LSP decays into three jets with a
lifetime about $\mathcal{O}(10^{-13}) \sim \mathcal{O}(10^{-12})$ seconds,
if other superpartners have masses $\lesssim 1$~TeV.
If the coupling is too small ($\lambda^{''} \lesssim 10^{-5}$), it leads to a displaced vertex of the neutralino decay, again making discrimination
from background easier. If the coupling is too large ($\lambda \gsim  0.01$), the single squark production and/or
the branching ratio of squarks decaying to two quarks  would become sizeable.
In this case, the searches for the squark resonance would have a great sensitivity to constrain the model.  
A similar class of models has already been studied in ref.~\cite{Butterworth:2009qa}, where the analysis was focused on exploiting substructure of high-$p_T$ jets originating from heavily boosted neutralinos. Here, we take another approach, where the $p_T$-requirements for jets are relaxed, but on the other hand, a large jet multiplicity is required. Same-sign dilepton signal in the $UDD$ model was also analyzed in ref.~\cite{Baer:1994zw} for SUSY searches at the Tevatron.

To study constraints on such SUSY models, we reinterpret the results 
of the ATLAS large jet multiplicity plus missing energy search~\cite{Aad:2012hm} and the CMS same-sign dilepton with jets plus missing 
energy search~\cite{:2012th}. The lightest neutralino decay 
produces many additional jets compared to the $R$-parity conserving case while the amount of
missing energy is reduced. However, the missing energy cut is still required
because of a large QCD backround. Nevertheless, the ATLAS search still has a sufficient sensitivity to the $R$-parity 
violating signal where missing energy can originate from decays of weak bosons appearing in the SUSY cascade 
chain, $W\to\ell\nu, Z\to\nu \bar \nu$. Furthermore, an additional missing energy contribution will come from jet energy mismeasurement. 
The decays of weak bosons will also give charged leptons in the final state. Thus, the CMS same-sign 
dilepton search can also be expected to have sensitivity to the $R$-parity 
violating SUSY. Thus, the weak bosons in cascade decay chains play a key role in 
constraining the $R$-parity violating SUSY models using the current LHC 
analyses.

In the remaining of this section, we demonstrate that elektroweak bosons could be 
frequently produced in ``vanilla'' SUSY models. 
There are two main sources of weak bosons in such SUSY models.
One of them originates from decays of charginos/neutralinos from squarks
and gluinos cascade decay chains and the other comes from top squark decays. 
Firstly, we discuss the weak 
boson production in squark cascade decay. If $m_{\tilde{q}} > M_{\tilde{W}} > M_{\tilde{B}}$, left-handed 
squarks decay dominantly into a chargino, followed by the chargino decay into a
neutralino and a $W$ boson thanks to the higgsino admixture. On the other hand, if 
$m_{\tilde{q}} > M_{\tilde{B}} > M_{\tilde{W}}$, right-handed 
squarks decays are a source of the weak bosons. The right-handed squarks 
decay into the second lightest neutralino, which then decays to a chargino and a $W$ 
boson. In either case, final state leptons would become soft if we would abandon the GUT relation and make the bino and wino 
nearly mass-degenerate. However, if the $\mu$ parameter is of the similar order, 
the full chargino/neutralino sector would not be compressed due to the off-diagonal components of 
the gaugino-higgsino mass matrix. A mixing in the neutralino and chargino sectors would still result 
in a rich phenomenology with many decay chains possible. The $\mu$ term of the order of the electroweak scale is plausible 
in the context of ``natural'' SUSY, see e.g.~\cite{Kitano:2006gv,Asano:2010ut,Papucci:2011wy}. 
Finally, if $M_{\tilde{W}} >m_{\tilde{q}} > M_{\tilde{B}}$, squarks mainly 
decay into the lightest neutralino with one jet. Even in this case, stops (produced either directly or in gluino decay chains) 
could be a source of $W$ bosons originating from top quark decays. Stop tends to be the lightest squark because of a 
large top Yukawa coupling. The light stop is also motivated by naturalness.

If squarks are heavier than gluino, gluino decays via off-shell squarks to a pair of jets and a chargino or a neutralino. 
Thus, $W$ bosons are also produced from such a decay chain, as in the case 
of squark decays discussed above. If only stop is lighter than 
gluino, then gluino dominantly 
decays into stop and a top quark. Decays of left- and 
right-handed stop could produce $W$ bosons via top or gaugino decays.

Finally, we comment on the possibility of the SUSY decay chain including 
sleptons. If sleptons are light and SUSY particles decay into LSP with intermediate
sleptons, many leptons would be expected in the final state. This would result in a 
high sensitivity also in other search channels, see e.g.~\cite{Allanach:2001xz,Allanach:2001if,Chatrchyan:2012ye}.


\section{Signal regions and event generation  \label{sec:3}}

In this section we define signal regions that have already been used by ATLAS and CMS experiments
for SUSY searches. Furthermore, we provide details of event generation that was applied for our 
benchmark models in section~\ref{sec:4}.

\subsection{Signal regions   \label{subsec:1}}

As discussed in section~\ref{sec:2}, the $UDD$ type RPV SUSY models predict a large 
hadronic activity in the final state as the LSPs decay hadronically. Therefore, we expect that the ATLAS large jet multiplicity
plus missing energy search~\cite{Aad:2012hm} could provide strong constraints on such models.
In the CMS same-sign dilepton with jets plus missing energy search~\cite{:2012th}, 
the requirement on the missing energy is relaxed compared to the other 
SUSY searches, since the SM background can be well suppressed by requiring 
the same-sign dilepton and high-$p_T$ jets. Because the RPV models predicts low amounts of missing energy, this search should 
offer good sensitivity to such models. Three signal regions are chosen 
from each of the original ATLAS and CMS searches and these are summarised in tables~\ref{tab:ATSR} and~\ref{tab:CMSR}, respectively.\footnote{
We have checked that the other signal regions are less sensitive to our RPV SUSY models.
}

The ATLAS search defines the signal regions according to the number of jets with $p_T > 55$~GeV.
The three signal regions demand at least 7, 8 and 9 jets, respectively.
In each case an event is vetoed if it contains isolated leptons with $p_T > 20$ (10)~GeV for electrons (muons).
In addition, it is required that $E_T^{\rm miss}/\sqrt{H_T} > 4 \,{\rm GeV}^{1/2}$, where $H_T$ is the scalar sum of 
$p_T$ of jets with $p_T > 40$~GeV, $|\eta|<2.8$.
After these cuts, the SM background is well suppressed.
The main background comes from fully hadronic and semi-leptonic $t \bar t$ events
and QCD multi-jet events.  
The number of expected background events, observed events and the $95\%$ CL model-independent upper limit on the number of
BSM events for the three signal regions are listed in table~\ref{tab:ATSR}.

The three signal regions in the CMS search are classified based on the requirement on the minimum threshold of $E_T^{\rm miss}$:
$120$, $50$ and $0$~GeV.
The SM background is well suppressed mainly because of the requirements of at least one same-sign dilepton pair with $p_T$ greater than
20 (10)~GeV for the leading (subleading) lepton,  
at least two jets (with $p_T > 40$~GeV, $|\eta|<2.5$) 
and $H_T > 450$~GeV, where $H_T$ is calculated using jets with $p_T > 40$~GeV and $|\eta|<2.5$.
In order to suppress a soft lepton background originating from heavy hadron decays, a minimum dilepton invariant mass of 8~GeV is imposed.     
In the background, leptons come from heavy gauge boson decays, leptonic decays within jets
and jets mimicking leptons. 
The number of background events, observed events and the $95\%$ CL upper limit on the number of
BSM events are listed in table~\ref{tab:CMSR}.

\begin{table}
\begin{center}
\begin{tabular}{|c|c|c|c|}
\hline
signal region & 7j55 & 8j55 & 9j55 
\\ \hline \hline
$N_{\rm lepton} (p_T^{e, \mu} > 20, 10 \,{\rm GeV})$ & \multicolumn{3}{|c|}{ $= 0$ }\\ 
\hline
$E_{T}^{\rm miss}/\sqrt{H_T}$ & \multicolumn{3}{|c|}{ $ > 4 \,\sqrt{\rm GeV}$ }\\ 
\hline
$N_{\rm jet}(p_T>55 \,{\rm GeV})$ & $\ge 7$ & $\ge 8$ & $\ge 9$ \\
\hline  \hline
$N$(BG) & $167 \pm 34$ & $17 \pm 7$ & $1.9\pm0.8$ \\ 
\hline  
$N$(observed) & 154 & 22 & 3 \\ 
\hline  
$N_{\rm BSM}^{95\% {\rm UL}}$ & 64 & 20 & 5.7 \\ 
\hline  
\end{tabular} 
\caption{ The signal regions defined in the ATLAS large jet multiplicity plus missing energy search.  
The number of expected background events, observed events and the $95\%$ CL model-independent upper limit on
the number of BSM events for each signal region are also shown. 
For more details see~\cite{Aad:2012hm}. 
\label{tab:ATSR}
}
\end{center}
\end{table}

\begin{table}
\begin{center}
\begin{tabular}{|c|c|c|c|}
\hline
signal region & MET120 & MET50 & MET0 
\\ \hline \hline
$N(\rm SS~lepton~pair)$ & \multicolumn{3}{|c|}{ $\ge 1$ }\\ 
\hline
$N_{\rm jet}(p_T>40 \,{\rm GeV})$ & \multicolumn{3}{|c|}{ $ \ge 2$ } \\ 
\hline
$p_T^{ \ell 1, \ell 2}$ & \multicolumn{3}{|c|}{ $> 20, 10$\,{\rm GeV} }\\ 
\hline
$m(\ell^+_i \ell^-_i)$ & \multicolumn{3}{|c|}{ $ > 8 \,{\rm GeV}$ }\\ 
\hline
$H_T$ & \multicolumn{3}{|c|}{ $ > 450 \,{\rm GeV}$ }\\ 
\hline
$E_T^{\rm miss}$ & $> 120\,{\rm GeV}$ & $> 50\,{\rm GeV}$ & $> 0\,{\rm GeV}$ \\ 
\hline  \hline
$N$(BG) & $4.9 \pm 2.6$ & $13.0 \pm 4.9$ & $23.6\pm8.4$ \\ 
\hline  
$N$(observed) & 4 & 11 & 16 \\ 
\hline  
$N_{\rm BSM}^{95\%{\rm UL}}$ & 9.6 & 6.2 & 10.4 \\ 
\hline  
\end{tabular} 
\caption{ The signal regions defined in the CMS same-sign dilepton with jets plus missing energy search.  
The number of expected background events, observed events and the $95\%$ CL model-independent upper limit on
the BSM events for each signal region are also shown. 
For more details see~\cite{:2012th}.
\label{tab:CMSR}
}
\end{center}
\end{table}

\subsection{Event generation and detector simulation   \label{subsec:2}}

In our analysis, RPV SUSY events are generated using 
\texttt{Herwig++~2.5.2}~\cite{Bahr:2008pv,Gieseke:2011na,Gigg:2007cr}
with $\sqrt{s} = 7$~TeV.
The detector response is simulated using \texttt{Delphes~2.0.2}~\cite{Ovyn:2009tx} with a corresponding
detector card depending on the analysis.
Jets are clustered by the anti-$k_T$ algorithm with a radius parameter of 0.4 (0.5) in the ATLAS (CMS) analysis.

In the ATLAS analysis, electrons are required to have $p_T>20$~GeV and $|\eta|<2.47$, whilst muons should have
$p_T>10$~GeV and $|\eta|<2.4$.
The lepton isolation is checked in the following way.
First, any jet candidate lying within a distance $\Delta R = \sqrt{(\Delta \eta)^2   + (\Delta \phi)^2} = 0.2$
of an electron is discarded. 
A lepton is considered to be isolated if there are no other jets around the lepton candidate within a distance of $\Delta R = 0.4$.

In the CMS analysis, all lepton candidates must have $|\eta|<2.4$.
For the lepton isolation, a scalar sum of transverse track momenta (excluding the lepton track itself)
and transverse calorimeter energy deposits within $\Delta R < 0.3$ is computed.
If the sum is less than $15\%$ of the lepton $p_T$, the lepton is considered to be isolated.
In this study, the lepton isolation is important not only for the CMS same-sign dilepton analysis, but also for
the ATLAS large jet multiplicity search.
The ATLAS analysis requires some amount of missing energy, $E_T^{\rm miss}/\sqrt{H_T} > 4 \,{\rm GeV}^{1/2}$,
which may be achieved if the event contains neutrinos in the RPV SUSY scenario.
In general, neutrinos are produced from gauge boson or tau decays and are often associated with charged leptons.
Such events would, however, be rejected due to the lepton veto if the efficiency of the lepton acceptance was perfect.
 
Before deriving constraints on the RPV CMSSM, we validate
our event and detector simulation by reproducing the $R$-parity conserving (RPC) CMSSM exclusion limits 
in the ($m_0, m_{1/2}$) plane  
provided by ATLAS and CMS.
We found that a difference between our exclusion contours and the published ones was typically less than 50~GeV
in $m_{1/2}$. The results are therefore compatible within uncertainties estimated by ATLAS.
To obtain the exclusion limits, we calculate the acceptance times efficiency for different SUSY production processes
separately in each signal region
using the simulated events.
Then we calculate the next-to-leading order  cross sections using     
\texttt{Prospino~2.1}~\cite{prospino1, prospino2}. 
From those values, we estimate the number of expected signal events and compare it with the reported model-independent upper limit
on the number of BSM events. We find a good agreement between our result and those obtained by ATLAS and CMS.

\section{Constraining vanilla SUSY models with R-parity violation  \label{sec:4}}

In this section, we show that the ``vanilla'' SUSY models with $R$-parity 
violation can already be constrained by the current LHC searches. For this purpose, 
we reinterpret the following two direct SUSY 
searches: the ATLAS large jet multiplicity plus missing energy 
search~\cite{Aad:2012hm} and the CMS same-sign dilepton with jets plus 
missing energy search~\cite{:2012th}, that have been carried 
out using an integrated luminosity of 4.7 and 4.98~fb$^{-1}$ at 
$\sqrt{s}=7$~TeV, respectively. We show that these analyses have a good sensitivity 
for the $UDD$ type $R$-parity violation models. 
As sample 
spectra, we consider the CMSSM and a simplified model where the  
sleptons, higgsinos and third generation squarks are decoupled. 
In the latter case, missing energy and leptons in the final state come solely from decays of gauge bosons, as discussed in section~\ref{sec:3}.

\subsection{CMSSM + \texorpdfstring{${UDD}$}{uud} model  \label{subsec:3}}

Our first example for the RPV SUSY scenario is the CMSSM with the $U_2 D_1 D_2$ operator in the superpotential
and the small coupling, $\lambda^{\prime \prime}_{212} = 10^{-3}$, see eq.~\eqref{eq:UDD}.
This choice of the coupling does not alter the low energy mass spectrum compared to the RPC CMSSM.
Sparticle production cross sections and cascade decay chains remain identical, 
until the cascade decay chains reach the LSPs.
In the RPV CMSSM, the lightest neutralinos further decay into three quarks, eq.~\eqref{eq:decay}, increasing the hadronic activity in the final state,
whilst in the RPC CMSSM the neutralinos are stable and contribute to the transverse missing energy.
Furthermore, we fix $\tan\beta = 10$, $A_0=0$, $\mu > 0$ throughout the paper.\footnote{
If the trilinear coupling $A_0$ is large, e.g.\ to realize the $126$~GeV Higgs boson
\cite{:2012gk,:2012gu}, the stop mass would change. Since in this case the $\tilde{t}_1$ 
could become lighter and more abundant in gluino decay chain, our results would provide 
a conservative limit. 
The Higgs boson mass constraint can also be satisfied by extending the MSSM.
For example, in the NMSSM, $U(1)$-extended models and models with vector-like matter, 
the bounds can be satisfied without introducing the large $A$-term or heavy stops. 
Our result can be applied to those models, if the branching ratios relevant to our study, such as
$\tilde q \to \tilde \chi_1^{\pm} q \to \tilde \chi_1^0 W^{\pm} q$, do not change much, which is usually the case.
}
The low energy spectrum is calculated using \texttt{SOFTSUSY-3.2.4}~\cite{Allanach:2009bv}
and the decay branching ratios are obtained using \texttt{SUSYHIT}~\cite{Djouadi:2006bz}.\footnote{
We keep the Higgs
boson mass as calculated by \texttt{SOFTSUSY}, which is typically in the range 110-115~GeV. 
Taking a correct value $m_h = 126 \gev$ would affect our constraint by shifting the kinematical threshold
of the $\neu{2} \to h \neu{1}$ decay in the ($m_0, m_{1/2}$) plane by $\sim 50$~GeV in the $m_{1/2}$ direction.
Since the $\neu{2}$ production is subdominant compared to the $\cha{1}$ production 
across the ($m_0, m_{1/2}$) plane,
we expect this effect to be below the systematic uncertainty on the exclusion curves
($\sim 50$~GeV in the $m_{1/2}$ direction). In any case, this makes our limits slightly 
more conservative than with the correct Higgs mass.  
}

\begin{figure}[t]
\begin{center}
  \subfigure[]{\includegraphics[width=0.45\textwidth]{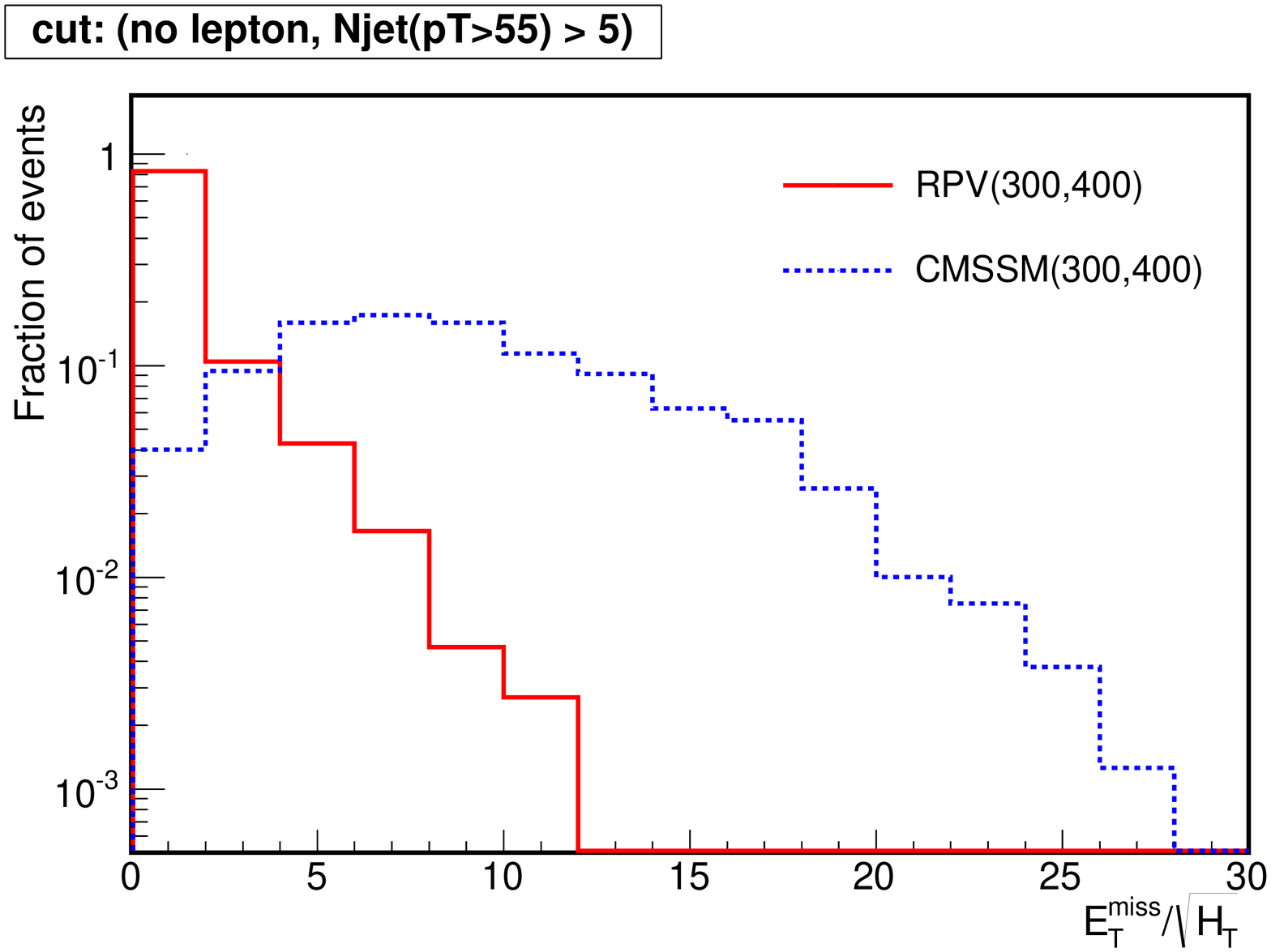}\label{fig:h_ymet}}
  \hspace{2mm}
  \subfigure[]{\includegraphics[width=0.45\textwidth]{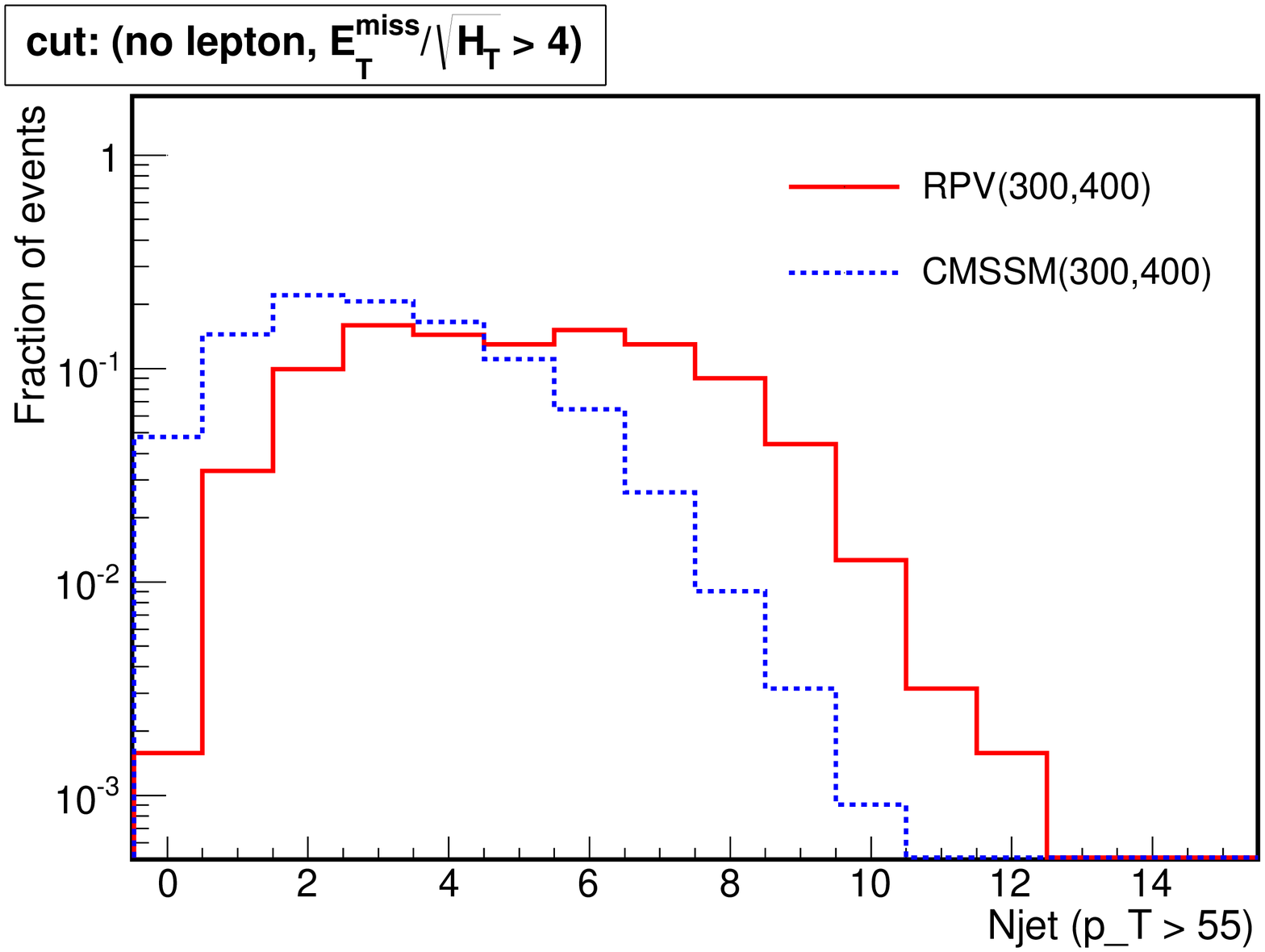}\label{fig:h_njets}}
\caption{ Distribution of (a) $E_T^{\rm miss}/\sqrt{H_T}$ for events with at least 6 jets of  
$p_T>55~{\rm GeV}$; and (b) the number of jets with $p_T>55~{\rm GeV}$ fulfilling 
$E_T^{\rm miss}/\sqrt{H_T} > 4~{\rm GeV}^{1/2}$. Lepton veto was applied in both cases. 
The red solid (blue dashed) histograms are for the RPV (RPC) CMSSM, respectively. \label{fig:njymet} }
\end{center}
\end{figure}

In figure~\ref{fig:h_ymet}, the red solid (blue dashed) histogram shows the distribution of $E_T^{\rm miss}/\sqrt{H_T}$
in the RPV (RPC) CMSSM, respectively.   
In this example, we have chosen $m_0 = 300$~GeV, $m_{1/2}=400$~GeV as a representative model point. 
The distributions are obtained after two pre-selection cuts:  no isolated lepton and 
at least 6 jets with $p_T>55~{\rm GeV}$. 
As can be seen, in the RPC CMSSM the distribution peaks around 7~${\rm GeV}^{1/2}$ and has a long tail.
On the other hand, in the RPV model, the distribution peaks at zero and quickly falls off towards higher values.
The reason is two fold. The missing transverse energy is significantly reduced in the RPV model
because neutralinos cannot contribute to $E_T^{\rm miss}$.
Secondly, the hadronic decays of neutralinos increase $H_T$, which further decreases $E_T^{\rm miss}/\sqrt{H_T}$.     
The number of events that pass the $E_T^{\rm miss}/\sqrt{H_T} > 4~{\rm GeV}^{1/2}$ cut in the RPV model 
is reduced by more than one order of magnitude compared to the RPC model.

Figure~\ref{fig:h_njets} shows the distribution of the number of jets with $p_T > 55$~GeV
after lepton veto and the $E_T^{\rm miss}/\sqrt{H_T} > 4~{\rm GeV}^{1/2}$ cut.
As can be seen, the number of jets is enhanced in the RPV model because of the hadronic decays of the neutralinos.
In the 8- and 9-jets bins, the enhancement is by factor of 10.

\begin{figure}[t]
\begin{center}
  \subfigure[]{\includegraphics[width=0.48\textwidth]{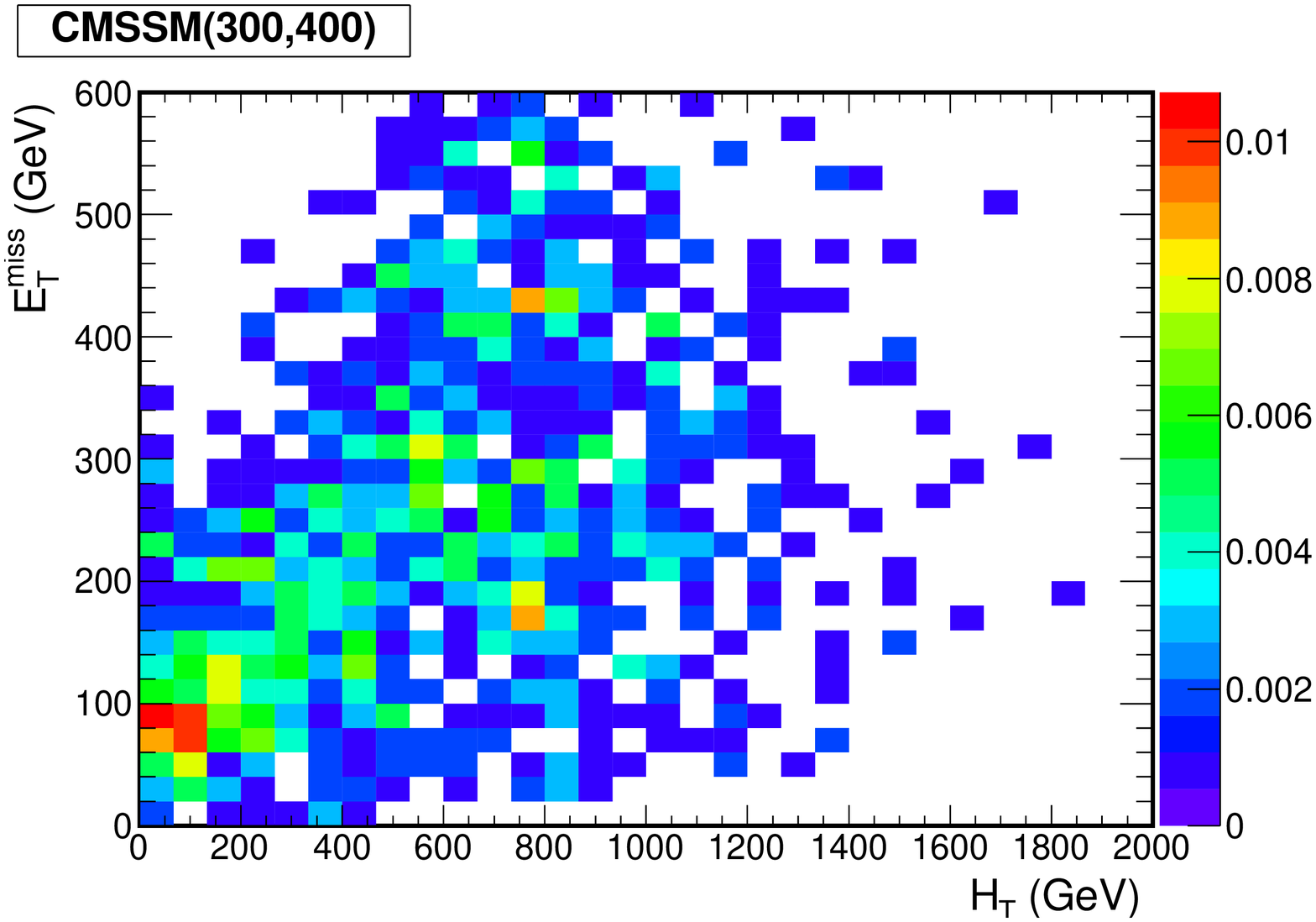}\label{fig:htvsmet_cm}} \hspace{0.2cm}
  \subfigure[]{\includegraphics[width=0.48\textwidth]{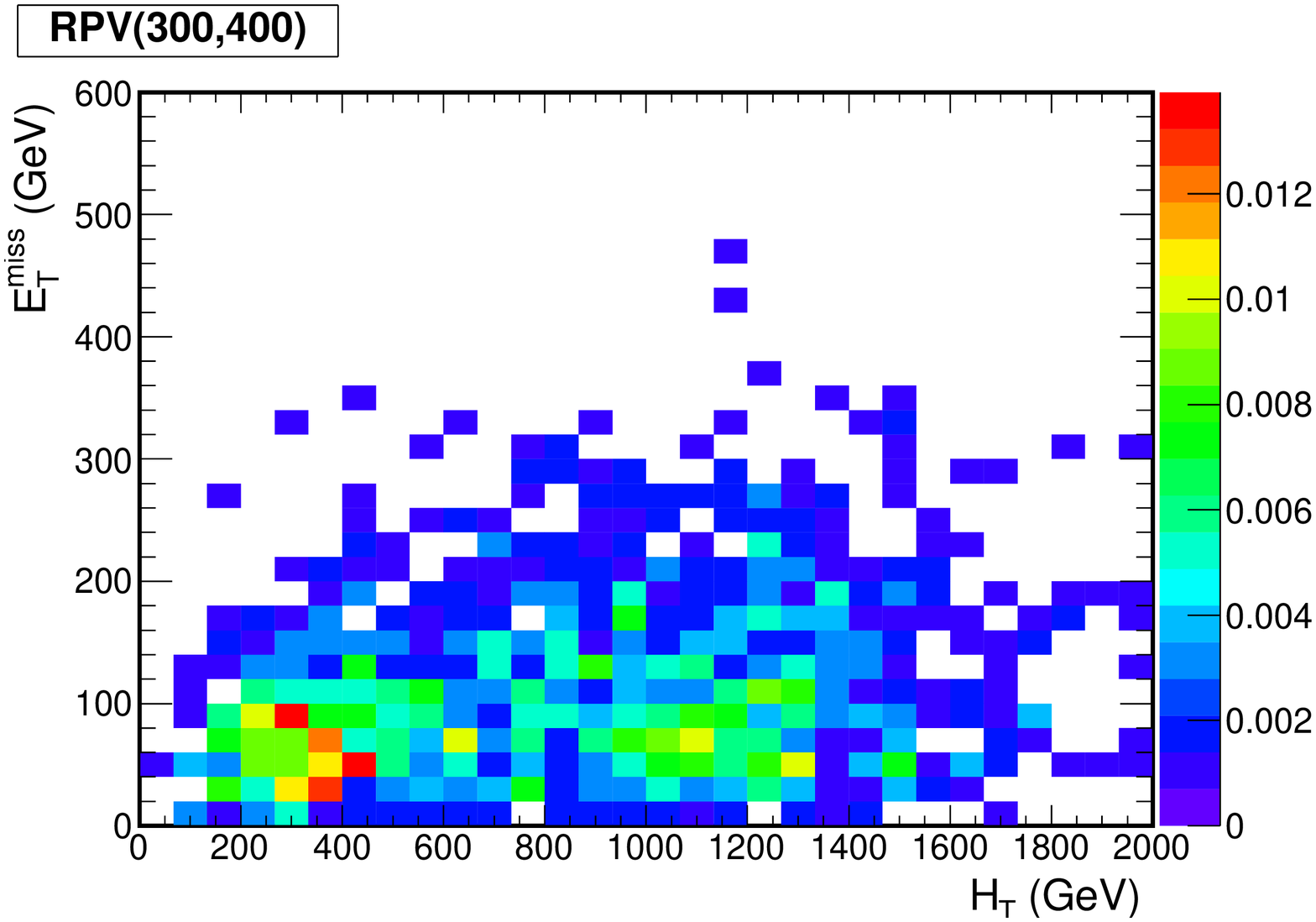}\label{fig:htvsmet_rpv}}
\caption{ Event density distributions in the ($H_T$, $E_T^{\rm miss}$) plane in (a) the RPC CMSSM and (b) the RPV CMSSM. \label{fig:htvsmet} }
\end{center}
\end{figure}

Figure~\ref{fig:htvsmet} shows the event density distributions in the ($H_T$, $E_T^{\rm miss}$) plane.
The left, \ref{fig:htvsmet_cm}, and the right, \ref{fig:htvsmet_rpv}, panels correspond to the RPC and RPV models, respectively.   
In the RPC model, the events are more scattered over the plane.
A large proportion of the events fall into the $E_T^{\rm miss} > 200$~GeV region, while the density decreases
if $H_T$ exceeds 900~GeV.
On the other hand, in the RPV model, the events are more confined in the $E_T^{\rm miss} < 200$~GeV region,
whilst the event density does not decrease until $H_T$ reaches 1400~GeV.

\begin{figure}[t]
\begin{center}
  \includegraphics[width=0.75\textwidth]{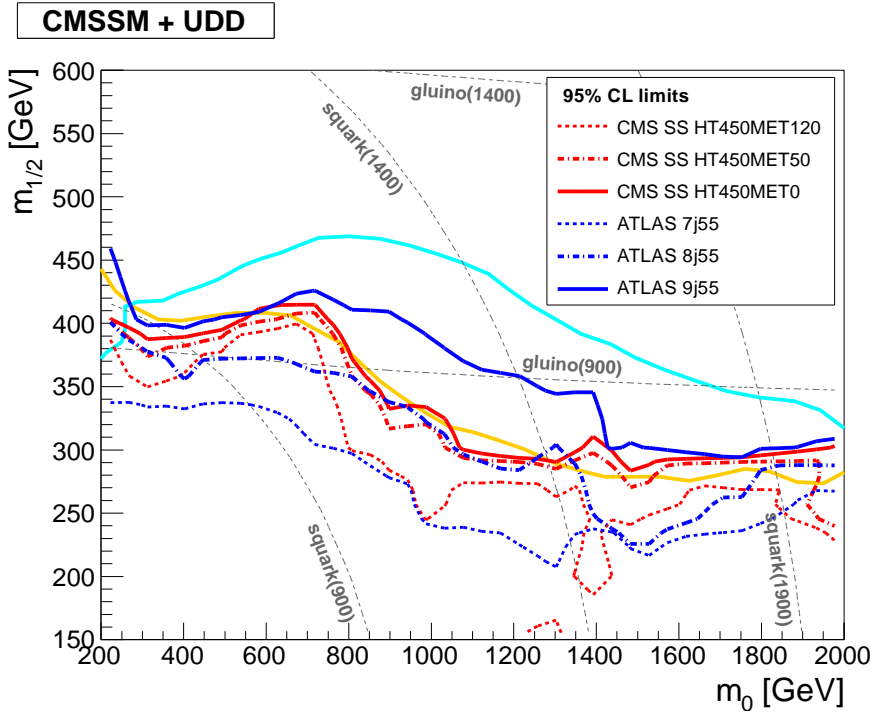}
\caption{The exclusion limits in the $R$-parity violating CMSSM from six ATLAS (blue lines) and CMS (red lines) signal regions defined in section~\ref{subsec:1}.  The original CMSSM exclusion contours obtained from ATLAS (cyan curve)~\cite{Aad:2012hm} and CMS (orange curve)~\cite{:2012th} are also shown.    \label{fig:excl_cr} }
\end{center}
\end{figure}

Figure~\ref{fig:excl_cr} shows the $95\%$ CL exclusion limits in the RPV CMSSM parameter plane.
The original exclusion curves for the RPC CMSSM are shown as well for comparison.
In the ATLAS large jet multiplicity search, the 9j55 signal region places the strongest bound among the three signal regions, 
which was also the case for the RPC CMSSM.
The exclusions are slightly degraded in the large $m_0$ region because $E_T^{\rm miss}/\sqrt{H_T}$ is much smaller in the RPV case than
in the RPC model.
On the other hand, in the small $m_0$ region, the bound is even stronger than in the RPC CMSSM.
In the RPC CMSSM, obtaining a large jet multiplicity is quite difficult in the small $m_0$ region compared to 
the large $m_0$ region.
This is because in the small $m_0$ region, squarks decay into a neutralino or a chargino plus one jet in a two-body decay.
On the other hand, in the large $m_0$ region, squarks decay into a gluino plus one jet and the gluino 
further decays into a neutralino or a chargino plus two jets through a three-body decay, 
producing two additional jets in one cascade decay chain compared to the decay chain in the small $m_0$ region.
The RPV helps in this situation:   the hadronic decays of the LSPs provide several extra jets and 
make it easier to satisfy the 9-jet requirement even in the small $m_0$ region.

In the CMS same-sign dilepton search, the MET0 signal region puts the strongest bound. 
The sensitivities among the signal regions are reversed compared to the RPC model, i.e. 
the most constraining signal region in the RPC case gives the weakest constraint in the RPV case. 
This is expected because the signal regions with a harder $E_T^{\rm miss}$ cut lose more signal events in the RPV SUSY compared to the RPC case.
In this search, the exclusion limit is almost the same as in the RPC case.

In conclusion, those two searches can provide good constraints in the CMSSM type SUSY spectrum even when the $R$-parity is violated
and the LSPs decay hadronically.
For the equal gluino and squark masses, the searches exclude gluinos (squarks) up to 900~GeV.
The gluino mass limit does not depend strongly on the squark mass.
A 700~GeV gluino is excluded for any squark mass.

\subsection{Simplified spectrum + \texorpdfstring{$UDD$}{udd} model  \label{subsec:4}}

\begin{figure}[t]
\begin{center}
 {\includegraphics[width=0.75\textwidth]{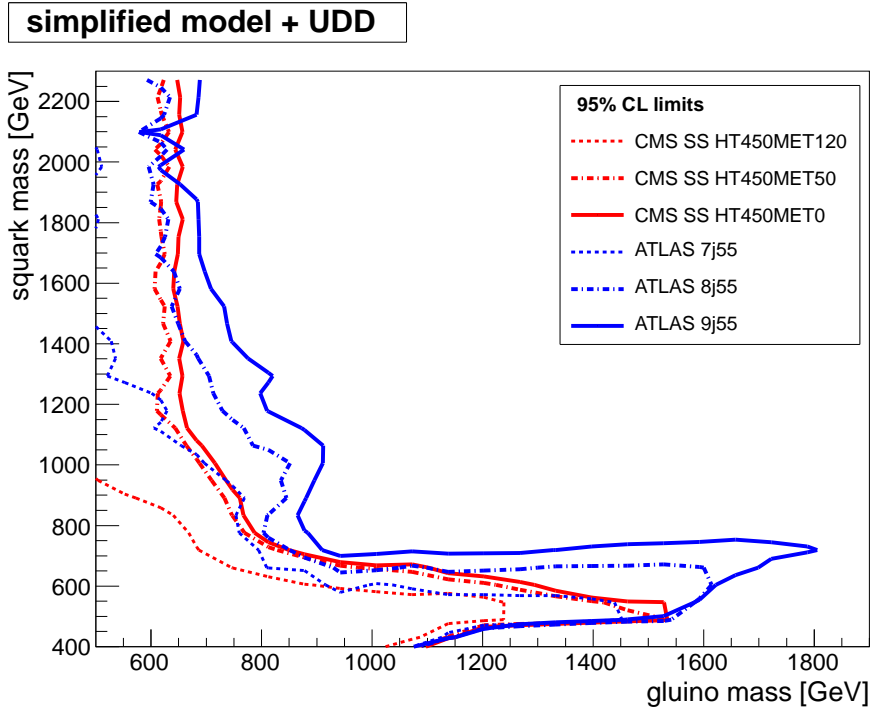}}
\caption{The exclusion limits in the $R$-parity violating simplified model from six ATLAS (blue lines) and CMS (red lines) signal regions defined in section~\ref{subsec:1}.  \label{fig:excl_simp} }
\end{center}
\end{figure}

Our second example is a simplified model with the $U_2 D_1 D_2$ operator.
In this model, the sleptons, higgsinos and third generation squarks are decoupled.
The first two generations of squarks are mass degenerate.
In the gaugino sector, an approximate GUT relation, $m_{\tilde g}: m_{\tilde W}: m_{\tilde B} = 6:2:1$, is imposed.
Therefore, there are only two free parameters,  which are relevant for the collider phenomenology:
the gluino and squark masses.
Decay branching ratios of SUSY particles are calculated using \texttt{SUSYHIT}~\cite{Djouadi:2006bz}.

Figure~\ref{fig:excl_simp} shows the exclusion limits in this RPV simplified model.
The features are similar to the RPV CMSSM results.
The ATLAS large jet multiplicity search puts a slightly better constraint than the CMS same-sign dilepton search.
The 9j55 and MET0 signal regions are the most sensitive ones in the ATLAS and CMS analyses, respectively.
If the squark and gluino masses are equal, the exclusion limit is 800~GeV. 
The gluino exclusion limits reach as far as $m_{\tilde{g}}=1800\gev$ for squark masses $m_{\tilde{q}}=700\gev$. 
We note however that lighter squarks are not
excluded in the high gluino mass region. 
This is because with the heavier gluino the wino also becomes heavy, exceeding 
the squark mass. 
In this case, the only allowed decay mode of squarks is $\tilde{q}\to q \neu{1}$, which does not
produce missing energy and leptons. 
One can see that the exclusion curve roughly follows the $m_{\tilde{g}}/3$ line. 
On the other hand, the gluino exclusion bound will not be strongly affected by squark masses in the heavy squarks  
limit. 
The gluino mass below 650~GeV is excluded independently of the squark mass.

The exclusion limits on the squark and gluino masses are slightly weaker than those obtained in RPV CMSSM.
This is because the events in the RPV CMSSM can contain several tops.
For example, in the CMSSM, stop is the lightest squark throughout most of the parameter space.
Therefore, gluino may preferentially decay to a stop plus a top if $m_{\tilde g} > m_{\tilde t} + m_t$
and to $b t \tilde \chi_1^{\pm}$ or $t t \tilde \chi_{1/2}^{0}$ via a three-body decay.
These tops increase the number of $W$s and jets in the final state.
On the other hand, in the simplified model, the sources of the same-sign dileptons and missing energy are mainly the decays of
$SU(2)$ gauginos into $W$ followed by the leptonic decay $W \to \ell \nu_{\ell}$.

\section{Summary and conclusions \label{summary}}

In this study, we have investigated constraints on SUSY models with $R$-parity violation 
from ATLAS and CMS experiments. In particular, we have focused on the $R$-parity 
violation by $UDD$ term as it appears to be potentially 
the most difficult to be searched at hadron colliders. If the LSP is 
the lightest neutralino, it decays into three jets in such models. In this case, the LSP decay
does not produce charged leptons or missing energy, which provide a
powerful discrimination against SM background in the current SUSY 
searches at the LHC. 

We point out that the weak bosons which are produced in the SUSY cascade
decay chains are a good source of charged leptons and missing 
energy. We show that the gluino and squarks lighter than a TeV can be 
already excluded in the RPV CMSSM if their masses are approximately equal. This means 
that the current LHC searches have sufficient sensitivity even for the $UDD$ 
type $R$-parity violating SUSY models. Therefore, 
a significant part of the SUSY parameter space which provides the electroweak symmetry 
breaking without fine-tuning have already been excluded also in the $R$-parity 
violating case.

Our results imply that $R$-parity violation could help in relaxing the bounds on the CMSSM. 
In our benchmark scenario we obtain the limit of $\sim 900$~GeV for squark and gluino masses in 
contrast to $\sim 1.4$~TeV in the $R$-parity conserving CMSSM. Searches in the final states of 
large jet multiplicity and same-sign dileptons seem to be well-suited for the $UDD$ RPV, replacing usual large missing energy searches. 
However, the constraints derived in this paper strongly depend on the abundance of charged leptons and neutrinos 
in the final states.
For models which do not expect leptons in the events, these constraints would be much weaker.
For example, in models where 
squarks and gluino can directly decay only into the lighest neutralino associated with one or two jets 
our bound is not applicable.
In this class of models, naturalness can nevertheless be achieved together with $R$-parity violation.

\subsection*{Note added}
While completing this paper an updated ATLAS study on large jet multiplicity final states has been published~\cite{ATLAS-CONF-2012-103}. We do not include those results in the present paper.

\acknowledgments
M.A. acknowledges support from the German Research Foundation (DFG) through
grant BR 3954/1-1.

\bibliographystyle{JHEP}
\bibliography{rpv_excl}

\end{document}